\documentclass[12pt]{iopart}
\usepackage{graphicx,graphics,color,epsfig}
\usepackage{iopams}

\begin{document}

\title[Multi-impurity polarons in a dilute Bose-Einstein condensate]{%
Multi-impurity polarons in a dilute Bose-Einstein condensate}

\author{D. H. Santamore\dag\ and Eddy Timmermans\ddag}

\address{\dag\ Department of Physics, Temple University, Philadelphia, PA 19122, USA}

\address{\ddag\ T-4, Theory Division, Los Alamos National Laboratory, Los Alamos, NM 87545, USA}

\begin{abstract}
We describe the ground state of a large, dilute, neutral atom Bose-Einstein
condensate (BEC) doped with $N$ strongly coupled mutually indistinguishable,
bosonic neutral atoms (referred to as \textquotedblleft impurity") in the
polaron regime where the BEC density response to the impurity atoms remains
significantly smaller than the average density of the surrounding BEC. We
find that $N$ impurity atoms with $N\neq1$ can self-localize at a lower
value of the impurity-boson interaction strength than a single impurity
atom. When the `bare' short-range impurity-impurity repulsion does not play
a significant role, the self-localization of multiple bosonic impurity atoms
into the same single particle orbital (which we call co-self-localization)
is the nucleation process of the phase separation transition. When the
short-range impurity-impurity repulsion successfully competes with
co-self-localization, the system may form a stable liquid of self-localized
single impurity polarons.
\end{abstract}

%\pacs{03.75.Hh,03.75.Mn,05.30.Rt,05.30.Jp}
%\submitto{\NJP}
\maketitle

\section{Introduction}

Cold atom traps offer intriguing advantages for studying strong interaction
physics in quantum many-body systems \cite{BDZ08}. We focus on the cold atom
prospect for conducting strongly coupled polaron studies. The polarons we
consider are neutral atoms (referred to as \textquotedblleft
impurity atoms\textquotedblright\ in this paper) distinguishable from the BEC atoms, i.e.,
atoms in a different spin state or atoms of a different species embedded in a
large dilute gas Bose-Einstein condensate (BEC). The BEC polaron is the impurity atom 
accompanied by its phonon cloud created by the interaction of the impurity
with the BEC atoms. By varying an external magnetic field near a specific
Feshbach resonant value \cite{KetterleN98, TTHK99, CGJT10}
experimentalists can tune the strength of the impurity-boson interactions and
bring the polaron into the strongly coupled regime. In that regime, the
BEC-polaron self-localizes \cite{CT06,KB06}: the ground state
properties become consistent with an impurity wavefunction of finite extent.
The recently demonstrated species-specific potentials (e.g. the impurity
atoms), such as laser generated potentials experienced by one kind of atoms
\cite{CBLRTMSI09,LCBNIM10}, provides unparalleled prospects for the
manipulation of such polaron objects.

Attractive impurity-boson interactions give a collapsed ground state in 2D and
3D \cite{BBJ08}. The cold atom realization of self-localized polarons with
repulsive impurity-boson interactions can demonstrate polaron physics in
regimes that are difficult to access in condensed matter \cite{Vbook}. Such
cold atom experiments can then address fundamental questions pertaining to the
nature of self-localization, the description of heavy polarons with a mass
that is comparable to the energy scale of the boson modes and the many-body
structure of systems with many polarons. At sufficiently low temperatures and
sufficiently high polaron densities, cold atom traps may realize polaron
superfluids without the complications that arise in describing the inevitable
Coulomb interactions in electronic polarons \cite{AM94}.

Arguably, the polaron system provides the paradigm of a strongly interacting 
many-body environment modifying the properties of a single particle \cite{Vbook}.
In the original polaron problem -- the
description of an electron in an ionic crystal -- Landau \cite{L33} and later,
Landau and Pekar \cite{LP48} showed that the electron self-localizes in the
strong coupling regime. In an attempt to understand the role of backflow in
the roton feature of the helium superfluid dispersion, Feynman and Cohen
studied a different polaron: an impurity atom in a condensed $^{4}$He
superfluid \cite{FC56}. The paper by Pines, Miller and Nozieres \cite{MPN62}
explicitly referred to this system as a polaron and pointed out the similarity
between Feynman's ansatz wavefunction and the wavefunction implied by a
second-order perturbation calculation. In this sense, the polarons here can be
defined as the excitations created through the interaction of the impurity with the BEC atoms.
Cold atom physics is now poised to
realize polarons of this type at strong coupling, creating structures that are
smaller than the coherence length of the BEC. These may provide a new handle
with sub-coherence length resolution.

The description of impurity-BEC polarons brings in aspects of related
problems. Self-localization occurs at intermediate coupling, which is the
challenging regime to describe. A Feynman path integral description
\cite{TCOKTD09} of the BEC-impurity polaron showed a surprising, non-monotonic
dependence of the polaron size with coupling strength near the coupling
strength at which the strong coupling description predicts self-localization
\cite{TCOKTD09N}. In this paper, we will work within the strong coupling
description in the assumption that this description correctly captures the
important trends. We describe the co-self-localization of multiple impurity
atoms in which the many-body ground state properties become consistent with
$N$ bosonic impurity atoms occupying the same single particle wavefunction of
finite extent. The starting point of \cite{CT06} and \cite{KB06}, the
assumption of a product state, was designed for the description of electron
bubbles in helium superfluids \cite{GR74}. Below, we show that the
linearization of the BEC response to the impurity wavefunction reproduces the
results of the Landau-Pekar elimination of the boson modes which couple as in
the Frohlich Hamiltonian to the single impurity atom in the Bogoliubov
approximation of the surrounding BEC. That description shows that the energy
reduction produced by the BEC density response can be cast into the form of an
attractive self-interaction potential. The linearization also shows that the
Frohlich impurity-BEC coupling, the Landau-Pekar elimination and the resulting
self-mediated interaction fail when the BEC-density response becomes
comparable to the density of the surrounding BEC. When the BEC-density is not
affected much by the localized impurity, the structure of the impurity-BEC
system is that of a polaron. When the BEC density is expelled by the impurity,
the impurity becomes a `bubble' separated spatially from the BEC. The physics
of multiple, bosonic self-localized impurities is connected with the
phenomenon of zero-temperature Bose-Einstein condensate phase separation
\cite{CF78}, \cite{Kett99}, \cite{T98}, as we point out below. The response of
a dilute BEC to a localized object was studied in the context of a moving
object that is significantly smaller than the healing length of the BEC
\cite{AP04} and in the context of a trapped ion interacting with the BEC atoms
\cite{MPS05} via the polarization interaction of charged and polarizable
particles. The polarization potential can also give rise to self-localization
of the ion as shown in \cite{CTD10}.

In this paper, we describe the ground state of a large BEC with multiple,
strongly coupled, mutually indistinguishable, bosonic impurity atoms embedded
in a large dilute gas Bose-Einstein condensate. In the absence of the impurity
atoms we take the BEC to be homogeneous. The boson-boson, boson-impurity and
impurity-impurity interactions are described by repulsive short-range
effective interactions (contact interactions). As mentioned, we assume that
the impurity atoms co-self-localize: the self-localized impurity atoms do not
only overlap, they occupy the same single-particle orbital. We find that
multiple impurity atoms can self-localize at smaller values of the
impurity-boson coupling constant, than a single impurity atom can. We identify
the minimal number of impurity atoms needed to stabilize the self-localized
polaron structure for fixed coupling constant as a property of the droplet
formation process in the nucleation of the BEC phase separation transition.
When the short-range impurity-impurity interaction is sufficiently large, its
competition with the self-localizing BEC-mediated interaction can lead to a
regime in which one impurity atom self-localizes but not two or more.

The paper is organized as follows: in section \ref{Sec_single_impurity}, we
discuss the magnitudes, dimensions and coupling constants of single
BEC-impurity self-localization. In section \ref{Sec_LandauPekar}, 
we use the Landau-Pekar method for describing the single neutral atom BEC-impurity problem so that
we can later proceed beyond the homogeneous phonon description and investigate
multiple impurities in the strong coupling limit. In section
\ref{Sec_multiimpurity}, we examine the system of strongly coupled
multi-impurity boson particles, followed by section \ref{Sec_analysis}, where
we examine the multi-impurity polaron structure variationally and discuss
the connection with BEC-phase separation dynamics. Finally, in section
\ref{Sec_conclusion} we conclude.

\section{Single impurity self-localization in BEC\label{Sec_single_impurity}}

First, we consider the zero-temperature self-localization of a neutral
impurity atom embedded in a dilute gas BEC. The impurity atom couples to the
elementary BEC excitations and forms a polaron. Strictly speaking, the
resulting object resembles a piezo-polaron since the BEC excitations are
acoustic phonon modes. The term `piezo-polaron' was originally used by Mahan
and Hopfield \cite{MH64},  \cite{M72book}, for conduction electrons
interacting with the acoustic modes in a piezoelectric crystal \cite{P69},
\cite{BGD7}. To bring the impurity atom to the strong coupling regime of
self-localization, we assume that either the impurity-boson scattering length
$a_{IB}$ is Feshbach tuned to a large positive value or the boson-boson
scattering length $a_{BB}$ is tuned to a small positive value. We describe the
orders of magnitude and discuss the experimental challenge of avoiding large BEC
depletion when impurity self-localization occurs. Since many BECs can be
described as locally homogeneous (i.e., they are in the Thomas-Fermi limit),
we consider the ground state of a homogeneous BEC of $N_{B}$ indistinguishable
boson atoms of mass $m_{B}$, confined to a macroscopic volume $\Omega$,
corresponding to an average density $\rho_{0}=N_{B}/\Omega$. Two bosons at
position $\mathbf{x}_{i}$ and $\mathbf{x}_{j}$ effectively interact via a
contact potential $v_{BB}(\mathbf{x}_{i}-\mathbf{x}_{j})=\left[  4\pi\hbar
^{2}a_{BB}/m_{B}\right]  \delta\left(  \mathbf{x}_{i}-\mathbf{x}_{j}\right)
$. The BEC is dilute, so that the gas parameter $\gamma=\sqrt{\rho_{0}%
a_{BB}^{3}}$ is much smaller than unity, i.e., $\gamma\ll1$. The chemical
potential of the BEC bosons is $\mu_{B}=\lambda_{BB}\rho_{0}$ and the BEC's
healing length - the distance over which the BEC-density tends to its
asymptotic value if the condensate vanishes at a planar boundary - follows
from equating the corresponding kinetic energy to $\mu_{B}$. We will work with
a coherence length $\xi$ for which $\hbar^{2}/[2m_{B}\xi^{2}]=2\mu_{B}$, which
gives $\xi=\sqrt{16\pi\rho_{0}a_{BB}}$. The excitation of a collective phonon
mode of momentum $q$ then costs energy
\begin{equation}
\hbar\omega_{\mathbf{q}}=\hbar\left\vert q\right\vert c\sqrt{1+\left(
\left\vert q\right\vert \xi\right)  ^{2}}\label{dispersion}%
\label{one}
\end{equation}
where $c=\sqrt{\mu_{B}/m_{B}}$ denotes the BEC sound velocity.

Throughout the paper, we refer to the BEC atoms as \textquotedblleft
B\textquotedblright\ and to the impurity neutral atoms as \textquotedblleft
I\textquotedblright\
. When impurity atoms of mass $m_{I}$ are added to the
BEC, the impurity atom at $\mathbf{r}_{i}$ interacts with a BEC atom at
$\mathbf{r}_{j}$ as well as with another impurity atom at $\mathbf{r}_{l}$ via
short-range interaction potentials $v_{IB}\left(  \mathbf{r}_{i}%
-\mathbf{r}_{j}\right)  =\lambda_{IB}\delta\left(  \mathbf{r}_{i}%
-\mathbf{r}_{j}\right)  $ and $v_{II}\left(  \mathbf{r}_{i}-\mathbf{r}%
_{l}\right)  =\lambda_{II}\delta\left(  \mathbf{r}_{i}-\mathbf{r}_{l}\right)
$, where $\lambda_{IB}=2\pi\hbar^{2}a_{IB}\left(  m_{I}^{-1}+m_{B}%
^{-1}\right)  $ and $\lambda_{II}=(4\pi\hbar^{2}a_{II}/m_{I})$ and
$a_{IB},a_{II}$ are the scattering lengths describing the low energy impurity-boson 
and boson-boson scattering processes.
We take all inter-particle interactions to be repulsive, i.e.,
$a_{IB},a_{BB},a_{II}>0$, where $a_{BB}$ is the scattering length of the
potential  $\lambda_{BB}=(4\pi\hbar^{2}a_{BB}/m_{B})$. If the impurities are
bosonic and coexist with the BEC-bosons at sufficient density and at
sufficiently low temperature, the system is a mixture of a \textquotedblleft
B\textquotedblright\ and an \textquotedblleft I\textquotedblright\ condensate.
The \textquotedblleft B\textquotedblright-BEC mediates interactions between
the impurity atoms that are described by an attractive Yukawa potential of
range $\xi$
\begin{equation}
V_{s}(\mathbf{r}_{i}-\mathbf{r}_{j})=-\frac{Q^{2}}{\left\vert \mathbf{r}%
_{i}-\mathbf{r}_{j}\right\vert }e^{-\left(  \left\vert \mathbf{r}%
_{i}-\mathbf{r}_{j}\right\vert /\xi\right)  }\label{Yukawa potential}%
\end{equation}
where we characterize the strength by an effective charge $Q$ to emphasize the
useful analogy with Coulomb interactions. The appearance of a Yukawa interaction
is not surprising. It is known in quantum field theory \cite{PSbook} that the interactions mediated by a scalar boson field (provided by the BEC) takes the form of an attractive Yukawa potential in a non-relativistic case. In accordance, the impurity atoms in
the \textquotedblleft I\textquotedblright-BEC distributed at density $\rho
_{I}\left(  \mathbf{r}\right)  $ experience a local chemical potential
$\mu_{I}\left(  r\right)  $ with a $\rho_{I}$ dependent contribution
\begin{eqnarray}
\mu_{I}\left(  \mathbf{r}\right)    & =\lambda_{IB}\rho_{0}+
\lambda_{II}\rho_{I}\left(  \mathbf{r}\right)  %\nonumber\\
-\int d^{3}\mathbf{r}^{\prime}\left[  Q^{2}/\left\vert \mathbf{r}%
-\mathbf{r}^{\prime}\right\vert \right]  e^{-\left\vert \mathbf{r}%
-\mathbf{r}^{\prime}\right\vert /\xi}\rho_{I}\left(  \mathbf{r}^{\prime
}\right)  \nonumber\\
& =\lambda_{IB}\rho_{0}+\left(  \lambda_{II}-4\pi\xi^{2}Q^{2}\right)  \rho_{I}%
\end{eqnarray}
if the impurity BEC is distributed homogeneously.

The above chemical potential term implies a diverging \textquotedblleft
I\textquotedblright\ compressibility ($\sim\lbrack\partial\mu_{I}/\partial
\rho_{I}]^{-1}$) when $\lambda_{II}\rightarrow4\pi\xi^{2}Q^{2}$. It is the
diverging compressibility that causes the zero-temperature BEC phase
separation that we describe in section \ref{Sec_analysis}. This transition
occurs when the Fetter-Colson relation \cite{CF78} is satisfied $\lambda
_{II}\rightarrow\lambda_{IB}\left(  \lambda_{IB}/\lambda_{BB}\right)  $, where
$\lambda_{BB}=(4\pi\hbar^{2}a_{BB}/m_{B})$ so that
\begin{equation}
Q^{2}=\lambda_{IB}\frac{\lambda_{IB}}{\lambda_{BB}}\frac{1}{4\pi\xi^{2}}%
=4\mu_{B}a_{BB}\left(  \frac{\lambda_{IB}}{\lambda_{BB}}\right)
^{2}\label{Q2}%
\end{equation}
is the value of the Yukawa strength parameter.

The BEC mediated interaction is a consequence of BEC density correlations: an
impurity atom $i$ at $\mathbf{r}_{i}$ causes a density variation in the BEC that
is experienced by impurity atom $j$ at $\mathbf{r}_{j}$ as a change in its BEC
mean field energy. Describing the impurity atoms by a quantum wavefunction, we
find that even a single impurity is affected by BEC density correlations. The
impurity interaction energy density at $\mathbf{r}$ is influenced by the
impurity density at $\mathbf{r}^{\prime}$ as the BEC responds to the density
of the entire impurity wavefunction. As we will see below, a single impurity
then experiences an effective self-interaction $V_{s}\left(  \mathbf{r}%
-\mathbf{r}^{\prime}\right)  =-$ $Q^{2}e^{-\left\vert \mathbf{r}%
-\mathbf{r}^{\prime}\right\vert /\xi}/\left\vert \mathbf{r}-\mathbf{r}%
^{\prime}\right\vert $. The integration of $V_{s}\left(  \mathbf{r}%
-\mathbf{r}^{\prime}\right)  $\ over $\mathbf{r}$ and $\mathbf{r}^{\prime}$
weighted by $\frac{1}{2}\rho_{I}(\mathbf{r})\rho_{I}(\mathbf{r}^{\prime})$
gives the gain in impurity-boson interaction energy by the adjustment of the
BEC density to the impurity wavefunction.

In the strong coupling limit, the effective self-interaction exceeds the
kinetic energy cost of localizing the impurity. For the Yukawa
self-interaction to overcome the kinetic energy, the impurity has to localize
to a size comparable to or less than $\xi$. In that case, the self-interaction
is Coulomb-like in most of the impurity-region and, hence, is efficient at
binding. As in the Hydrogen atom description, the length scale $R_{0}$\ is
determined by comparing Coulomb and kinetic energies
\begin{equation}
\frac{Q^{2}}{R_{0}}=\frac{\hbar^{2}}{m_{I}R_{0}^{2}}%
\end{equation}
Using equation (\ref{Q2}), $\mu_{B}=\hbar^{2}/[4m_{B}\xi^{2}]$, and $\xi
^{-2}=16\pi\rho_{0}a_{BB}$, we find that the effective Rydberg length is equal
to
\begin{equation}
R_{0}=\frac{1}{4\pi\rho_{0}a_{IB}^{2}}\frac{1}{\left(  1+\frac{m_{I}}{m_{B}%
}\right)  \left(  1+\frac{m_{B}}{m_{I}}\right)  }\label{Rydberg Length R0}%
\end{equation}
Aside from the mass factor $\left(  1+\frac{m_{I}}{m_{B}}\right)  ^{-1}\left(
1+\frac{m_{B}}{m_{I}}\right)  ^{-1}$, $R_{0}$ is the mean free path of an
impurity atom that encounters hard-sphere scatterers of radius $a_{IB}$
distributed at average density $\rho_{0}$. The energy scale of impurity
self-localization is given by the corresponding Rydberg energy
\begin{equation}
E_{0}=\frac{Q^{2}}{2R_{0}}=2\frac{m_{B}}{m_{I}}\mu_{B}\pi\rho_{0}\frac
{a_{IB}^{4}}{a_{BB}}\left(  1+\frac{m_{I}}{m_{B}}\right)  ^{2}\left(
1+\frac{m_{B}}{m_{I}}\right)  ^{2}.\label{Rydberg Energy E0}%
\end{equation}
The mediated self-interaction can only bind with sufficient efficiency if the
natural impurity size $R_{0}$\ is comparable to or shorter than $\xi$. The ratio of the BEC-coherence and Rydberg lengths, $\xi/R_{0}$, is the only parameter left in the description of the impurity wavefunction after scaling the energy and length by $E_{0}$ and $R_{0}$, respectively. We call this dimensionless ratio, the impurity-boson coupling parameter,
\begin{equation}
\beta\equiv\frac{\xi}{R_{0}}=\sqrt{\pi}\sqrt{\rho_{0}\frac{a_{IB}^{4}}{a_{BB}}%
}\left(  1+\frac{m_{I}}{m_{B}}\right)  \left(  1+\frac{m_{B}}{m_{I}}\right).
\end{equation}
Calculations \cite{CT06} indicate that a single impurity atom can
self-localize if $\beta>\beta_{c}\left(  1\right)  =4.7$, where $\beta
_{c}\left(  1\right)  $ stands for the critical $\beta$ for a single impurity atom.
As $\beta$ further increases, the self-localized impurity size shrinks to become a
point-like object (smaller than the BEC coherence length). The calculation of
a point-like potential moving with velocity $v$ \cite{AP04} gave an energy
$m_{D}v^{2}/2$ with $m_{D}=\frac{2}{3}m_{B}\sqrt{\pi}\sqrt{\frac{\rho
_{0}a_{IB}^{4}}{a_{BB}}}$, so that we might expect the effective polaron mass
$m^{\ast}$ to be equal to $m^{\ast}=m_{I}+m_{D}$ or
\begin{equation}
\frac{m^{\ast}}{m_{I}}=1+\frac{2\beta}{3\left(  1+\frac{m_{I}}{m_{B}}\right)
^{2}}%
\end{equation}
However, the calculation in \cite{AP04} was based on perturbation theory so
that its validity does not extend to the large coupling limit. For the classic
polaron treatment that describes an electron coupled to optical phonons,
Feynman estimated an effective mass that varies as the fourth power of the
coupling constant \cite{Vbook}.

It is useful to express the observables in terms of the impurity-boson
coupling constant $\beta$. For instance,
\begin{eqnarray}
Q^{2}  &  =2\frac{m_{B}}{m_{I}}\beta^{2}\mu_{B}R_{0}\\
E_{0}  &  =2\frac{m_{B}}{m_{I}}\beta^{2}\mu_{B}%
\end{eqnarray}
with $\tau_{0}=\hbar/E_{0}$ representing the time scale relevant to impurity self-localization.

The polaron description of the BEC-impurity assumes that the phonon modes are
excitations of a homogeneous BEC. However, as the impurity-boson interaction
is increased to reach the critical coupling $\beta_{c}\left(  1\right)  $, the
impurity can deplete the \textquotedblleft B\textquotedblright\ condensate in
its vicinity. The above description breaks down when the local BEC-density
variation $\delta\rho_{B}$ becomes comparable to $\rho_{0}$. For repulsive
impurity-boson interactions we expect the single impurity wavefunction to take
on a phase-separated bubble profile when $\delta\rho_{B}\simeq\rho_{0}$. While
this is interesting by itself, the impurity loses its Frohlich-coupled polaron
character in the bubble limit. As the self-interaction energy is the change in
boson-impurity energy caused by the impurity induced variation in BEC-density,
$\lambda_{IB}\delta\rho_{B}\sim E_{0}$ or
\begin{equation}
\frac{\delta\rho_{B}}{\rho_{0}}\simeq\frac{E_{0}}{\lambda_{IB}\rho_{0}}%
=\frac{E_{0}}{\mu_{B}}\frac{\lambda_{BB}}{\lambda_{IB}}=4\pi^{1/4}\sqrt
{\frac{m_{B}}{m_{I}}}\beta^{\frac{3}{2}}\gamma^{\frac{1}{2}}%
\label{deltaRhoB self}%
\end{equation}
where $\gamma$ represents the BEC gas parameter, $\gamma=\sqrt{\rho_{0}a_{BB}^{3}}$, defined earlier in this section (above equation (\ref{one})).

Can the self-localization of a single BEC-impurity be realized in cold atom
experiments? At the end of the next section, we discuss whether cold atom
technology can access the polaron regime -- the parameter region where the
density variation, $\delta\rho_{B}$, in equation (\ref{deltaRhoB self}) remains significantly smaller than $\rho_{0}$. Here
we discuss the experimental accessibility of the strong impurity-boson
coupling regime, $\beta>5$. Three experimental challenges must be met. Two of
them relate to the requisite increase in impurity-boson scattering length.
First of all, if that increase is effected by a magnetically controlled
Feshbach resonance, the magnetic field has to be sufficiently homogeneous and
constant to avoid significant variations of the impurity-boson interactions.
Secondly, three body recombination into the large, two atom impurity-boson
bound (dimer) state that becomes degenerate with the incident channel
impurity-boson continuum as $a_{IB}$ diverges, should not occur before the BEC
impurity has self-localized. As a third condition, we mention that the
BEC-temperature should be sufficiently low to prevent thermal fluctuations
from de-self-localizing the strongly coupled BEC impurity.

How large does $a_{IB}$ have to be? Estimating the necessary $a_{IB}$ from%
\begin{equation}
\frac{a_{IB}}{a_{BB}}=\sqrt{\frac{\beta/\gamma}{\sqrt{\pi}\left(
1+\frac{m_{I}}{m_{B}}\right)  \left(  1+\frac{m_{B}}{m_{I}}\right)  }},
\end{equation}
with $\gamma\sim10^{-4}$ and a realistic mass ratio of $\sim10$, we find that
$\frac{a_{IB}}{a_{BB}}\sim10^{2}$ if $\beta\sim10$. The Feshbach scattering
length with magnetic field dependence is given by%
\begin{equation}
a_{IB}=a_{IB,0}\left[  1-\frac{\Delta}{B-B_{0}}\right]  ,
\end{equation}
where $a_{IB,0}$ denotes the background scattering length, $\Delta$ is the
resonance width, and $B_{0}$ is the resonant field strength. Hence, a slow
magnetic field variation $\delta B$ induces an $a_{IB}$ variation%
\begin{equation}
\delta a_{IB}=a_{IB,0}\frac{\Delta}{\left(  B-B_{0}\right)  ^{2}}\delta B.
\end{equation}
For $B$ near $B_{0}$, $a_{IB}\simeq-a_{IB,0}\Delta/\left(  B-B_{0}\right)  $, so
that%
\begin{equation}
\frac{\delta a_{IB}}{a_{IB}}=\frac{a_{IB}}{a_{IB,0}}\left\vert \frac{\delta
B}{\Delta}\right\vert .
\end{equation}
Requiring $\left\vert \delta a_{IB}/a_{IB}\right\vert <\varepsilon$ leads to
$\left\vert \delta B/\Delta\right\vert <\varepsilon\left(  a_{IB,0}%
/a_{IB}\right)  $. Assuming, for instance, $a_{IB,0}/a_{IB}=1/30$,
$\varepsilon\sim0.03$, we obtain $\left\vert \delta B/\Delta\right\vert
<10^{-3}$. For $\Delta=1G$, the magnetic field variation is $\delta B<1mG$,
which can be achieved in today's laboratories. In contrast, a broader resonance may
require $\delta B<10mG$ $\left(  \Delta=10G\right)  $ or $\delta B<0.1G$
$\left(  \Delta=100G\right)  $.

We now estimate the recombination time of the BEC impurity atom in the worst
case scenario. Applying the same reasoning as in \cite{FRS96} where the
three-body recombination rate for a BEC of large scattering length was first
calculated, we expect a recombination rate of the form%
\begin{equation}
\frac{1}{\tau_{R}}\simeq\frac{3.9}{\sqrt{3}}\sqrt{1+2\left(  1+\frac{m_{B}%
}{m_{I}}\right)  }\frac{\hbar}{m_{B}}a_{IB}^{4}\rho_{0}^{2}.
\label{recombination}%
\end{equation}
For a BEC of large and positive $a_{BB}$, substituting $m_{I}\rightarrow m_{B}$ 
and $a_{IB}\rightarrow a_{BB}$ in equation (\ref{recombination}) gives an overestimate
of the recombination rate (see \cite{EGB99}). The authors of reference \cite{EGB99} 
found that the $a^{4}$ power has to be multiplied by an oscillating function of 
magnitude less than unity. Near a node of the oscillating function, the recombination 
rate is much reduced. Nevertheless, using equation (\ref{recombination}), we find%
\begin{equation}
\tau_{R}=\frac{\tau_{B}}{\beta}\frac{\left(  1+\frac{m_{B}}{m_{I}}\right)
^{2}\left(  1+\frac{m_{I}}{m_{B}}\right)  ^{2}}{\sqrt{1+2\frac{m_{B}}{m_{I}}}%
}\frac{4\pi^{2}\sqrt{3}}{3.9},
\end{equation}
where $\tau_{B}=\hbar/\mu_{B}$ represents the time scale on which the BEC can
respond to the impurity and form the phonon cloud that accompanies the impurity
in the polaron state. Note that for $m_{I}\sim10m_{B}$, $\beta\sim10$,
$\tau_{R}$\ can exceed $\tau_{B}$\ by more than two orders of magnitude.
Finally, we comment on the temperature requirement for observing BEC impurity
self-localization. Determining the polaron properties in a finite temperature
Feynman path integral calculation, reference \cite{TCOKTD09} identified polaron
self-localization with a sudden and unexpected increase in polaron size as
$\beta$ increases near the critical value $\beta_{c}$ obtained from the strong
coupling description. The polaron size increase was observed for a
$^{6}Li$-impurity in a $^{23}Na$-BEC in the case where the temperature
($k_{B}T$) was smaller than half the boson chemical potential. Therefore, we
expect that self-localization takes place when $k_{B}T<\mu_{B}f$, where $f$
can be as large as $0.5$. For a BEC of $\mu_{B}=\hbar\times300Hz$, the
temperature requirement of $T<7.5nK$ can be easily achieved in today's cold
atom experiments.

\section{Landau-Pekar elimination of phonon modes\label{Sec_LandauPekar}}

References \cite{CT06} and \cite{KB06} describe the self-localized
BEC-impurity state as a product wavefunction. Specifically, the many-body
state of an impurity of position $\mathbf{r}$ and $N_{B}$ bosons of position
$\mathbf{x}_{1},\mathbf{x}_{2},\ldots\mathbf{x}_{N_{B}}$, 
Bose-condensed in volume $\Omega$ so that the average boson density is
$\rho_{0}=N_{B}/\Omega$, was represented by a wavefunction $\psi\left(
\mathbf{r}\right)  \chi\left(  \mathbf{x}_{1}\right)  \chi_{2}\left(
\mathbf{x}_{2}\right)  \ldots\chi\left(  \mathbf{x}_{N}\right)  $, where
$\psi$ and $\chi$ denote the normalized single-particle impurity and BEC boson
wavefunctions, respectively. In the absence of an impurity, the BEC-field
$\varphi\left(  \mathbf{r}\right)  =\sqrt{N_{B}}\chi\left(  \mathbf{r}\right)
$ is $\varphi=\sqrt{\rho_{0}}$, position independent for a homogeneous BEC.
Assuming that the BEC response to the impurity remains smaller than
$\sqrt{\rho_{0}}$, $\varphi=\sqrt{\rho_{0}}+\delta\varphi$, the equations can be
linearized in $\delta\varphi$. The resulting description is equivalent to the
Landau-Pekar treatment we describe below. The product wavefunction reveals the
connection with the analogous \textquotedblleft fluid with embedded
droplet\textquotedblright\ system. The linearization procedure indicates how
to proceed beyond the homogeneous phonon description. On the other hand, the
Landau-Pekar method is most easily generalized to describe multiple impurities
in the strong coupling limit.

In the absence of an impurity, the excitations of the dilute BEC are
well-described in the Bogoliubov approximation. This description transforms
the operators that create and annihilate boson particles of momentum
$\mathbf{q}$, $\hat{b}_{\mathbf{q}}^{\dag}$, $\hat{b}_{\mathbf{q}}$ to
quasi-particle (phonon) creation and annihilation operators $\hat
{a}_{\mathbf{q}}^{\dag}$, $\hat{a}_{\mathbf{q}}$, that diagonalize the
linearized Hamiltonian. Adding a constant term, the transformed
BEC-Hamiltonian reads
\begin{equation}
\sum_{\mathbf{q}}\hbar\omega_{\mathbf{q}}\left(  \hat{a}_{\mathbf{q}}^{\dag
}\hat{a}_{\mathbf{q}}+\frac{1}{2}\right),
\end{equation}
where $\hbar\omega_{\mathbf{q}}$ is stated in equation (\ref{dispersion}). The
first step of the Bogoliubov procedure replaces the $\hat{b}_{\mathbf{q}%
=0}^{\dag}$, $\hat{b}_{\mathbf{q}=0}$-operators by $\sqrt{N_{B}}$ and expands
the energy in powers of $\sqrt{N_{B}}$. Accordingly, the boson density
operator becomes
\begin{eqnarray}
\hat{\rho}_{B,\mathbf{q}}  & \simeq N_{B}\delta_{\mathbf{q},0}+\sqrt{N_{B}%
}\left(  \hat{b}_{\mathbf{q}}^{\dag}+\hat{b}_{\mathbf{q}}\right)  \nonumber\\
& =N_{B}\delta_{\mathbf{q},0}+\sqrt{N_{B}}\sqrt{\frac{(\hbar^{2}q^{2}/2m_{B}%
)}{\hbar\omega_{\mathbf{q}}}}\left(  \hat{a}_{\mathbf{q}}^{\dag}+\hat
{a}_{\mathbf{q}}\right)
\end{eqnarray}
where the last step followed from implementing the Bogoliubov transformation.

Introducing the impurity density operator $\hat{\rho}_{I,\mathbf{q}}$, the
impurity-boson interaction takes the form
\begin{equation}
\frac{\lambda_{IB}}{\Omega}\sum_{\mathbf{q}}\hat{\rho}_{I,-\mathbf{q}}%
\hat{\rho}_{B,\mathbf{q}}\simeq\lambda_{IB}\rho_{0}\hat{n}_{I}+\frac
{\lambda_{IB}\sqrt{\rho_{0}}}{\sqrt{\Omega}}\sum_{\mathbf{q}}\sqrt
{\frac{\left(  \hbar^{2}q^{2}/2m_{B}\right)  }{\hbar\omega_{\mathbf{q}}}}%
\hat{\rho}_{I,-\mathbf{q}}\left(  \hat{a}_{\mathbf{q}}^{\dag}+\hat
{a}_{-\mathbf{q}}\right)  ,
\end{equation}
where $\hat{n}_{I}$ represents the number operator of the I-atoms, $\hat{\rho
}_{I,\mathbf{k}=0}=\hat{n}_{I}$. The second term on the right hand side
resembles the electron-phonon interaction term in the Frohlich Hamiltonian
\begin{equation}
\frac{M_{0}}{\sqrt{\Omega}}\sum_{\mathbf{q}}\nu_{-\mathbf{q}}\hat{\rho
}_{I,-\mathbf{q}}\left(  \hat{a}_{\mathbf{q}}^{\dag}+\hat{a}_{-\mathbf{q}%
}\right)  \label{Froh}%
\end{equation}
where $M_{0}=\lambda_{IB}\sqrt{\rho_{0}}$ and $\nu_{-\mathbf{q}}%
=\nu_{\left\vert \mathbf{q}\right\vert }=\sqrt{\left[  \left(  \hbar^{2}%
q^{2}/2m_{B}\right)  /\hbar\omega_{\mathbf{q}}\right]  }$. The phonon part of
the impurity-BEC Hamiltonian, $\hat{H}_{p}$, reads
\[
\hat{H}_{p}=\sum_{\mathbf{q}}\left[  \frac{\hbar\omega_{q}}{2}\left(  \hat
{a}_{\mathbf{q}}^{\dagger}\hat{a}_{\mathbf{q}}+\frac{1}{2}\right)
+\frac{\sqrt{2}M_{0}}{\sqrt{\Omega}}\nu_{-\mathbf{q}}\hat{\rho}_{I,-\mathbf{q}%
}\left(  \hat{a}_{\mathbf{q}}^{\dagger}+\hat{a}_{-\mathbf{q}}\right)  \right]
.
\]
The strong coupling approximation assumes that the impurity is described by a
well-defined wavefunction that is not entangled with the boson B state. As a
consequence, the coupling to other impurity states can be neglected and the
impurity density operator $\hat{\rho}_{I,\mathbf{q}}$ can be replaced by its
expectation value, $\hat{\rho}_{I,\mathbf{q}}\rightarrow\rho_{I,\mathbf{q}}$
$=\int d^{3}r\exp(-i\mathbf{q}\cdot\mathbf{r})\left\vert \psi\left(
\mathbf{r}\right)  \right\vert ^{2}$. To bring out the oscillator nature of
the boson modes, we introduce the oscillator coordinate and momentum operators
that are equivalent to the creation and annihilation operators,
\begin{eqnarray}
\hat{\phi}_{\mathbf{q}}  & =\frac{\hat{a}_{\mathbf{q}}^{\dagger}+\hat
{a}_{-\mathbf{q}}}{\sqrt{2}}\nonumber\\
\hat{\pi}_{\mathbf{q}}  & =\frac{\hat{a}_{\mathbf{q}}^{\dagger}-\hat
{a}_{-\mathbf{q}}}{i\sqrt{2}}\;.
\end{eqnarray}
In this notation, the phonon Hamiltonian $\hat{H}_{p}$ becomes a sum over
second order polynomials in $\phi$,
\begin{equation}
\hat{H}_{p}=\sum_{\mathbf{q}}\left[  \frac{\hbar\omega_{\mathbf{q}}}{2}\left(
\hat{\phi}_{\mathbf{q}}^{\dag}\hat{\phi}_{\mathbf{q}}+\hat{\pi}_{\mathbf{q}%
}^{\dag}\hat{\pi}_{\mathbf{q}}\right)  +\frac{\sqrt{2}M_{0}}{\sqrt{\Omega}}%
\nu_{-\mathbf{q}}\rho_{I,-\mathbf{q}}\hat{\phi}_{\mathbf{q}}\right]
.\label{Hp1}%
\end{equation}
Completing the squares in equation (\ref{Hp1}), we write the phonon Hamiltonian as
a sum over displaced oscillator Hamiltonians
\begin{equation}
\hat{H}_{p}=\sum_{\mathbf{q}}\frac{\hbar\omega_{\mathbf{q}}}{2}\left(
\hat{\Phi}_{\mathbf{q}}^{\dag}\hat{\Phi}_{\mathbf{q}}+\hat{\pi}_{\mathbf{q}%
}^{\dag}\hat{\pi}_{\mathbf{q}}\right)  -\sum_{\mathbf{q}}\frac{M_{0}^{2}%
}{\Omega}\frac{\nu_{\mathbf{q}}^{2}}{\hbar\omega_{\mathbf{q}}}\rho
_{I,\mathbf{q}}\rho_{I,-\mathbf{q}}\label{Hp}%
\end{equation}
where the $\hat{\Phi}$-operator denote the displaced boson oscillator coordinates
\begin{equation}
\hat{\Phi}_{\mathbf{q}}=\hat{\phi}_{\mathbf{q}}+\frac{\sqrt{2}M_{0}}%
{\sqrt{\Omega}}\frac{\nu_{\mathbf{q}}\rho_{I,\mathbf{q}}}{\hbar\omega
_{\mathbf{q}}}.
\end{equation}
As the displaced ground state energy is the same as that of the original
oscillator, the energy reduction caused by the phonon response to the impurity
density is given by the remainder
\begin{equation}
\Delta E_{p}=-\sum_{\mathbf{q}}\frac{M_{0}^{2}}{\Omega}\frac{\nu_{\mathbf{q}%
}^{2}}{\hbar\omega_{\mathbf{q}}}\rho_{I,\mathbf{q}}\rho_{I,-\mathbf{q}}.
\end{equation}
Using $\rho_{I,\mathbf{q}}=\int d^{3}\mathbf{r}\exp\left[  i\mathbf{q\cdot
r}\right]  \left\vert \psi\left(  \mathbf{r}\right)  \right\vert ^{2}$ this
expression can be cast in the form of an interaction energy
\begin{equation}
\Delta E_{p}=\frac{1}{2}\int d^{3}\mathbf{r}\int d^{3}\mathbf{r}^{\prime}%
|\psi\left(  \mathbf{r}\right)  |^{2}V_{s}\left(  \mathbf{r}-\mathbf{r}%
^{\prime}\right)  |\psi\left(  \mathbf{r}^{\prime}\right)  |^{2},
\end{equation}
where $V_{s}$ represents the inverse Fourier transform of $-2M_{0}^{2}%
\nu_{\mathbf{q}}^{2}/\hbar\omega_{q}$. The expression holds regardless of the
momentum dependence of $\omega_{\mathbf{q}}$ and $\nu_{\mathbf{q}}$ which are
different for ionic crystal (optical) and piezoelectric (acoustic) polarons.
In the case of the BEC-impurity or \textquotedblleft Bogoliubov
polaron\textquotedblright, $V_{s}$ is the attractive Yukawa interaction,
\begin{eqnarray}
V_{s}\left(  \mathbf{r}\right)    & =\frac{1}{\left(  2\pi\right)  ^{3}}\int
d^{3}\mathbf{q}\exp\left[  -i\mathbf{q\cdot r}\right]  \left(  \frac
{-2M_{0}^{2}\nu_{\mathbf{q}}^{2}}{\hbar\omega_{\mathbf{q}}}\right)
\nonumber\\
& =-\lambda_{IB}\frac{\lambda_{IB}}{\lambda_{BB}}\frac{1}{4\pi\xi^{2}}\frac
{1}{\left(  2\pi\right)  ^{3}}\int d^{3}qe^{(-i\mathbf{q}\cdot\mathbf{r}%
)}\left(  \frac{4\pi}{\xi^{-2}+q^{2}}\right)  \nonumber\\
& =-Q^{2}\frac{e^{-\left\vert \mathbf{r}\right\vert }}{\left\vert
\mathbf{r}\right\vert }%
\end{eqnarray}
where we used equation (\ref{Q2}) to obtain the above expression. The total change
in ground state energy $E$ caused by doping the BEC with a strongly coupled
impurity atom, includes the kinetic energy cost of localizing the impurity is
\begin{eqnarray}
E  & =\lambda_{IB}\rho_{0}\int d^{3}\mathbf{r}\;\left\vert \psi\left(
\mathbf{r}\right)  \right\vert ^{2}+\int d^{3}\mathbf{r}\;\psi^{\ast}\left(
\mathbf{r}\right)  \left(  \frac{-\hbar^{2}\nabla^{2}}{2m_{I}}\right)
\psi\left(  \mathbf{r}\right)  \nonumber\\
& +\frac{1}{2}\int d^{3}\mathbf{r}\int d^{3}\mathbf{r}^{\prime}\left\vert
\psi\left(  \mathbf{r}\right)  \right\vert ^{2}\left(  -Q^{2}\frac
{e^{-\left\vert \mathbf{r-r}^{\prime}\right\vert /\xi}}{\left\vert
\mathbf{r-r}^{\prime}\right\vert }\right)  \left\vert \psi\left(
\mathbf{r}^{\prime}\right)  \right\vert ^{2}\;.\label{energy E}%
\end{eqnarray}
The optimal wavefunction $\psi$ follows from minimizing $E$ while ensuring
that $\psi\left(  \mathbf{r}\right)  $ is normalized. We include the
normalization constraint by introducing a Lagrange multiplier $\mu$. The
extremum of the associated functional $F=E-\mu\left[  \int d^{3}%
\mathbf{r}\left\vert \psi\left(  \mathbf{r}\right)  \right\vert ^{2}-1\right]
$ is found by requiring $\delta F/\delta\psi^{\ast}\left(  \mathbf{r}\right)
=0$, which leads to
\begin{eqnarray}
\mu_{I}\psi\left(  \mathbf{r}\right)    & =-\frac{\hbar^{2}\nabla^{2}}{2m_{I}%
}\psi\left(  \mathbf{r}\right)  \nonumber\\
& +\int d^{3}\mathbf{r}^{\prime}\left(  -\frac{Q^{2}e^{-\left\vert
\mathbf{r}-\mathbf{r}^{\prime}\right\vert /\xi}}{\left\vert \mathbf{r}%
-\mathbf{r}^{\prime}\right\vert }\right)  \left\vert \psi\left(
\mathbf{r}^{\prime}\right)  \right\vert ^{2}\psi\left(  \mathbf{r}\right)
,\label{mu_IChi}%
\end{eqnarray}
where $\mu_{I}=\mu-\lambda_{IB}\rho_{0}$. The impurity wavefunction that
solves equation (\ref{mu_IChi}) is localized if and only if $\mu_{I}$\ is
negative. Multiplying equation (\ref{mu_IChi}) by $\psi^{\ast}\left(
\mathbf{r}\right)  $, integrating over $\mathbf{r}$,\ and making use of the
normalization condition, we obtain
\begin{eqnarray}
\mu_{I}  & =\int d^{3}\mathbf{r}\;\psi^{\ast}\left(  \mathbf{r}\right)
\left(  \frac{-\hbar^{2}\nabla^{2}}{2m_{I}}\right)  \psi\left(  \mathbf{r}%
\right)  \nonumber\\
& +\int d^{3}\mathbf{r}\int d^{3}\mathbf{r}^{\prime}\left\vert \psi\left(
\mathbf{r}\right)  \right\vert ^{2}\left(  -Q^{2}\frac{e^{-\left\vert
\mathbf{r-r}^{\prime}\right\vert /\xi}}{\left\vert \mathbf{r-r}^{\prime
}\right\vert }\right)  \left\vert \psi\left(  \mathbf{r}^{\prime}\right)
\right\vert ^{2},
\end{eqnarray}
which differs from equation (\ref{energy E}) by the factor of 2 in the
self-interaction term.

The numerical solution of equation (\ref{mu_IChi}), as calculated iteratively in
\cite{CT06}, shows that the profile of $\psi$ generally resembles a Gaussian
function. Assuming a normalized Gaussian wavefunction, $\psi\left(
\mathbf{r}\right)  =e^{-\left\vert \mathbf{r}\right\vert ^{2}/2\sigma^{2}%
}/\left(  \pi\sigma^{2}\right)  ^{3/4}$, we solve the problem variationally by
minimizing $E$ with respect to $\sigma$. To know if the variational
description predicts self-localization, we test if at the minimal width
$\bar{\sigma}$, $\mu_{I}\left(  \bar{\sigma}\right)  <0$. Working out the
integrals we find
\begin{equation}
E\left(  \sigma\right)  =-\frac{Q^{2}}{2\sigma}\sqrt{\frac{2}{\pi}}f\left(
\frac{\sigma}{\xi}\right)  +\frac{3}{4}\frac{\hbar^{2}}{m_{I}\sigma^{2}},
\end{equation}
where $f$,
\begin{eqnarray}
f\left(  y\right)    & \equiv& \int_{0}^{\infty}duu\exp\left[  -\frac{u^{2}}%
{2}\right]  \exp\left(  -uy\right)  \nonumber\\
& =& 1-\sqrt{\frac{\pi}{2}}y \exp\left[  -\frac{y^{2}}{2}\right]
\textrm{erfc} \left(  \frac{y}{\sqrt{2}}\right),
\end{eqnarray}%
accounts for the reduced efficiency of the Yukawa self-interaction when the
Gaussian width exceeds the BEC-coherence length (as $f(y)\rightarrow0$ when
$y\gg1$). If we scale the energies by the Rydberg energy of
equation (\ref{Rydberg Energy E0}), $E\rightarrow\left(  E/E_{0}\right)  $,
$\mu_{I}\rightarrow\left(  \mu_{I}/E_{0}\right)  $, and the length by the
Rydberg length of equation (\ref{Rydberg Length R0}), $x=\sigma/R_{0}$, $\beta
=\xi/R_{0}$, then $E$ and $\mu_{I}$ take the form
\begin{eqnarray}
\frac{E\left(  x\right)  }{E_{0}}  &  =-\sqrt{\frac{2}{\pi}}f\left(  \frac
{x}{\beta}\right)  \frac{1}{x}+\frac{3}{2x^{2}},
\label{Escaled}\\
\frac{\mu_{I}\left(  x\right)  }{E_{0}}  &  =-\sqrt{\frac{2}{\pi}}f\left(
\frac{x}{\beta}\right)  \frac{2}{x}+\frac{3}{2x^{2}}.
\end{eqnarray}

\begin{figure}[pth]
\begin{center}
\includegraphics[width=3in] {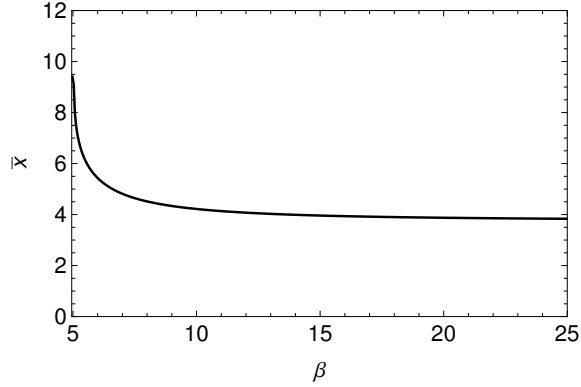}
\end{center}
\caption{Graph of the variationally determined Gaussian width of the single
impurity wavefunction in the strong coupling description of the impurity BEC
ground state, shown as a function of the boson-impurity coupling constant,
$\beta$.}%
\label{fig1a}%
\end{figure}

\begin{figure}[pth]
\begin{center}
\includegraphics[width=3.5in] {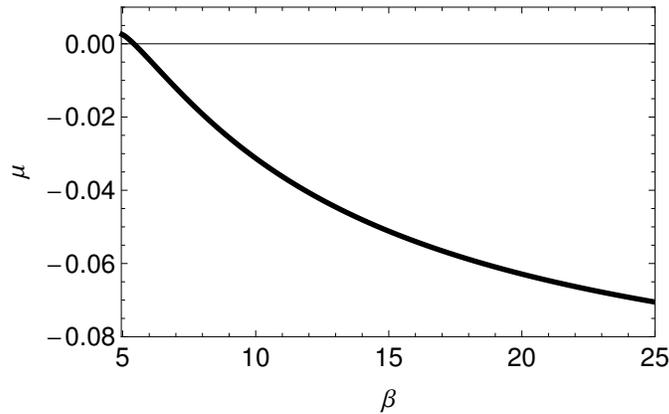}
\end{center}
\caption{The variationally determined chemical impurity potential $\mu_{I}$.
In the strong coupling description, the BEC-impurity polaron self-localizes
when $\mu_{I}<0$.}%
\label{fig1b}%
\end{figure}

In figure \ref{fig1a} we show the variationally determined Gaussian width
$\overline{\sigma}$ in units of the Rydberg length scale $R_{0}$. The extent
of the impurity wavefunction is proportional to this width with a factor of
proportionality that is $\sqrt{3/2}$, $\sqrt{<r^{2}>}=\sqrt{3/2}%
\overline{\sigma}$. In the Gaussian variational description, the bound state minimum of equation (\ref{Escaled}) disappears when $\beta<5$.
In figure \ref{fig1b}, we show the variationally obtained
$\mu_{I}$. As $\mu_{I}$ becomes negative when $\beta>5.4$, the Gaussian
variational description predicts that the BEC impurity polaron self-localizes when $\beta>\beta_{c}$ with $\beta_{c}=5.4$. Since $\beta_{c}$ gives the minimal coupling for self-localizing a single impurity atom, $N=1$, we will denote as $\beta_{c}(1)=5.4$.
%In comparison, the direct numerical solution of the non-linear Schrodinger %equation obtained in \cite{CT06} yielded $\beta_{c}=4.7$.
If the extent of the polaron is significantly smaller than the
BEC coherence length $\sigma\ll\xi$, equivalent to $x\ll\beta$, we can Taylor
expand $f$ as $f(x/\beta)\approx1-\sqrt{\pi/2}(x/\beta)$. In this limit, the
impurity system energy $E/E_{0}=-\sqrt{2/\pi}x^{-1}+3/\left[  2x^{2}\right]
+1/\beta$ attains a minimum at $\overline{x}=3\sqrt{\pi/2}$ corresponding to
$\sqrt{<r^{2}>}=(3/2)\sqrt{3\pi}R_{0}$ or $\overline{x}\rightarrow3\sqrt
{\pi/2}\approx3.76$, the large coupling limit that the graph of
figure \ref{fig1a} tends to as $\beta\rightarrow\infty$. Within the same
Taylor-expansion, $\mu_{I}/E_{0}=-\sqrt{2/\pi}(2/\overline{x})+(3/\left[
2\overline{x}^{2}\right]  )+2/\beta$, takes on a negative value if $\beta$
exceeds $2\pi$. Therefore, in the small polaron size approximation, $\beta
_{c}(1)\approx2\pi$, as compared to the variational result $\beta_{c}(1)=5.4$
and $\beta_{c}(1)=4.7$, numerically determined by solving the non-linear
equation, equation \ref{mu_IChi}.

The BEC density response to the self-localized impurity is a consequence of
the oscillator ($\phi$) displacement in the Landau-Pekar description. We
calculate the density variation in the Bogoliubov approximation,
\begin{eqnarray}
\delta\rho_{B}\left(  \mathbf{x}\right)    & =\rho_{B}\left(  \mathbf{x}%
\right)  -\rho_{0}\nonumber\\
& \simeq\frac{\sqrt{N_{B}}}{\Omega}\sum_{\mathbf{k}}e^{i\mathbf{k}%
\cdot\mathbf{x}}\left\langle \hat{b}_{\mathbf{k}}^{\dag}+\hat{b}_{-\mathbf{k}%
}\right\rangle \nonumber\\
& =\sqrt{\rho_{0}}\sqrt{\frac{2}{\Omega}}\sum_{\mathbf{k}}e^{i\mathbf{k}%
\cdot\mathbf{x}}\nu_{\mathbf{k}}\left\langle \hat{\Phi}_{\mathbf{k}}-\sqrt
{2}\frac{M_{0}}{\sqrt{\Omega}}\frac{\nu_{\mathbf{k}}}{\hbar\omega_{\mathbf{k}%
}}\rho_{I,\mathbf{k}}\right\rangle .
\end{eqnarray}
Using $\left\langle \hat{\Phi}_{\mathbf{k}}\right\rangle =0$ we find
\begin{equation}
\frac{\delta\rho_{B}\left(  \mathbf{x}\right)  }{\rho_{0}}=\frac{\sqrt
{\rho_{0}}}{M_{0}}\frac{1}{\Omega}\sum_{\mathbf{k}}e^{i\mathbf{k}%
\cdot\mathbf{x}}\left[  -\frac{2M_{0}^{2}\nu_{\mathbf{k}}^{2}}{\hbar
\omega_{\mathbf{k}}}\right]  \rho_{I,\mathbf{k}}.
\end{equation}
Recognizing the expression in the square brackets as the Fourier transform of
$V_{s}$ and replacing $\sqrt{\rho_{0}}/M_{0}=1/\lambda_{IB}$, we obtain
\begin{equation}
\delta\rho_{B}\left(  \mathbf{x}\right)  =-\frac{Q^{2}}{\lambda_{IB}}\int
d^{3}\mathbf{r}\frac{e^{-\left\vert \mathbf{x-r}\right\vert /\xi}}{\left\vert
\mathbf{x-r}\right\vert }\left\vert \psi\left(  \mathbf{r}\right)  \right\vert
^{2},
\end{equation}
which is identical to the expression found in \cite{CT06} by linearizing the
BEC-response. Note that the Landau-Pekar description and the linearization of
the product state give the same results for all observables.

We now determine the peak value of the BEC density variation. If we choose the origin at the center-of-mass of the impurity in $\left\vert \psi\left(  \mathbf{r}\right)  \right\vert ^{2}$, the peak density is reached at $\overline{x}=0$. In the true ground state, the impurity is not localized in any specific localization but the strong-coupling product wavefunction breaks translational symmetry. While the wavefunction does not have the correct symmetry, we know that some of the relevant polaron properties such as energy, mass, and polaron size give the correct values in the strong-coupling limit as they do in the traditional optical polaron problem. Assuming the width to take on the strong
coupling value, $\overline{\sigma}=3\sqrt{\pi/2}R_{0}$,
\begin{equation}
\left\vert \frac{\delta\rho_{B}\left(  \mathbf{x}=0\right)  }{\rho_{0}%
}\right\vert =\frac{E_{0}}{\mu_{B}}\frac{\lambda_{BB}}{\lambda_{IB}}%
\frac{4\sqrt{2}}{3\pi}f\left(  \frac{3\sqrt{\pi}}{2\beta}\right)
\end{equation}
where the factor $\frac{E_{0}}{\mu_{B}}\frac{\lambda_{BB}}{\lambda_{IB}}$ is
given by equation (\ref{deltaRhoB self}). Assuming $3\sqrt{\pi}/2\beta\ll1$, we
find
\begin{equation}
\left\vert \frac{\delta\rho_{b}\left(  \mathbf{x}=0\right)  }{\rho_{0}%
}\right\vert =\frac{16\sqrt{2}}{3\pi^{3/4}}\sqrt{\frac{m_{B}}{m_{I}}}%
\beta^{3/2}\gamma^{1/2}=3.196\sqrt{\frac{m_{B}}{m_{I}}}\beta^{3/2}\gamma
^{1/2}.
\end{equation}
In the polaron limit, the limit in which the Bogoliubov-approximated
impurity-boson coupling of the Frohlich type is valid, the BEC-density
response should be smaller than the average BEC density, $\delta\rho_{B}%
/\rho_{0}<\epsilon$, where $\epsilon$ is a fraction of unity. We find that the
average boson BEC-density should not exceed a value $\rho_{max}$ with
\begin{equation}
\rho_{max}=\frac{1}{a_{BB}^{3}}\left(  \frac{m_{B}}{m_{I}}\right)  ^{2}%
\frac{1}{\left(  3.196\right)  ^{4}}\frac{\epsilon^{4}}{\beta^{6}}%
=\frac{0.96\times10^{-2}}{a_{BB}^{3}}\left(  \frac{m_{B}}{m_{I}}\right)
^{2}\frac{\epsilon^{4}}{\beta^{6}}.
\end{equation}
Depending on the atoms trapped in the experiment, the above condition can pose
a challenge. Taking the impurity atom to be a spin-flipped atom in a $^{23}%
$Na-BEC, choosing $\epsilon=1/5$ and assuming that $a_{IB}$ is Feshbach tuned
while $a_{BB}=a_{Triplet}=3nm$, we find $\rho_{max}=3.6\times10^{10}cm^{-3}$,
which is a rather low density to make observations with. On the other hand,
embedding a $^{87}$Rb-impurity atom into $^{7}$Li BEC of $a_{BB}=0.5nm$, with
$\epsilon=1/5$ leads to $\rho_{max}=1.2\times10^{15}cm^{3}\left[  \beta
/\beta_{c}(1)\right]  ^{-6}$, yielding a maximal density of $2\times
10^{13}cm^{-3}$ even for $\beta=10$.

\section{Co-self-localization of bosonic
BEC-impurities\label{Sec_multiimpurity}}

Most schemes for realizing strong coupling
BEC-impurity polarons would create multiple impurities that are
indistinguishable in the same cold atom trap. In this section, we describe the
ground state of the $N$ impurity system in a homogeneous BEC. We assume that
the $N$ impurity atoms are indistinguishable bosons and that the
impurity-impurity interactions $\lambda_{II}\delta\left(  \mathbf{r}%
_{i}-\mathbf{r}_{j}\right)  $ are repulsive, $\lambda_{II}>0$.

We will characterize the strength of the short-range impurity-impurity
interactions by a third dimensionless coupling constant
\begin{equation}
\eta=\frac{\lambda_{II}R_{0}^{-3}}{\left(  2\pi\right)  ^{3/2}E_{0}}.
\end{equation}
The system's behavior is determined by three dimensionless parameters: the
BEC gas parameter $\gamma=\sqrt{\rho_{0}a_{BB}^{3}}$, which affects the
BEC-density response to the self-localized impurity but not the impurity
wavefunction in the polaron limit, the impurity-boson coupling constant,
$\beta=\sqrt{\pi}\sqrt{\rho_{0}a_{IB}^{4}/a_{BB}}\left(  1+m_{B}/m_{I}\right)
\left(  1+m_{I}/m_{B}\right)  $, and the above defined impurity-impurity
coupling constant, which can also be written as%
\begin{equation}
\eta=8\sqrt{2}\beta\gamma\left(  \frac{a_{II}}{a_{BB}}\right).
\label{eta}%
\end{equation}

We extend the above descriptions to treat $N$ impurities of position
$\mathbf{r}_{i}$, $i=1,2,\ldots N$, and $N_{B}$ bosons of position
$\mathbf{x}_{i}$, $i=1,2,\ldots N_{B}$. The strong coupling ground state
wavefunction can be written as a product state $\Psi\left(  \mathbf{r}%
_{1},\mathbf{r}_{2},\ldots\mathbf{r}_{N}\right)  \chi_{B}\left(
\mathbf{x}_{1}\right)  \chi_{B}\left(  \mathbf{x}_{2}\right)  \ldots\chi
_{B}\left(  \mathbf{r}_{N_{B}}\right)  $. Alternatively, the Landau-Pekar
treatment is easily generalized as the impurity-boson coupling is described by
the same Frohlich-like term of equation (\ref{Hp}). In the strong-coupling limit,
the density operator is replaced by its expectation value $\hat{\rho
}_{\mathbf{q}}\rightarrow\rho_{\mathbf{q}}$ where
\begin{equation}
\rho_{\mathbf{q}}=\int d^{3}\mathbf{r}_{1}\int d^{3}\mathbf{r}_{2}\ldots\int
d^{3}\mathbf{r}_{N}\ \sum_{j}e^{i\mathbf{q}\cdot\mathbf{r}_{j}}\left\vert
\psi\left(  \mathbf{r}_{1},\mathbf{r}_{2},\ldots\mathbf{r}_{N}\right)
\right\vert ^{2}.
\end{equation}
The strong coupling Landau-Pekar phonon elimination leads to a phonon-induced
ground state energy reduction
\begin{eqnarray}
\Delta E_{p}  &  =\frac{1}{2}\int d^{3}\mathbf{r}_{1}\int d^{3}\mathbf{r}%
_{2}\ldots\int d^{3}\mathbf{r}_{N}\ \int d^{3}\mathbf{r}_{1}^{\prime}\int
d^{3}\mathbf{r}_{2}^{\prime}\ldots\int d^{3}\mathbf{r}_{N}^{\prime}\nonumber\\
&  \times\ \left\vert \psi\left(  \mathbf{r}_{1},\mathbf{r}_{2},\ldots
\mathbf{r}_{N}\right)  \right\vert ^{2}\sum_{i,j}V_{s}\left(  \mathbf{r}%
_{i}-\mathbf{r}_{j}^{\prime}\right)  \left\vert \psi\left(  \mathbf{r}%
_{1}^{\prime},\mathbf{r}_{2}^{\prime},\ldots\mathbf{r}_{N}^{\prime}\right)
\right\vert ^{2}.
\end{eqnarray}
We can argue that the $i=j$ terms describe self-interaction while the $i\neq
j$ terms describe the BEC-mediated interactions. However, this distinction
becomes less meaningful when the impurities are indistinguishable and $\Psi$
satisfies the bosonic or fermionic permutation symmetry. In this case, all
$i,j$ contributions are equal and $\Delta E_{p}$ becomes
\begin{equation}
\Delta E_{p}=\frac{1}{2} N^{2}\int d^{3}\mathbf{r}\int d^{3}\mathbf{r}^{\prime}\rho
_{I}\left(  \mathbf{r}\right)  V_{s}\left(  \mathbf{r}-\mathbf{r}^{\prime
}\right)  \rho_{I}\left(  \mathbf{r}^{\prime}\right)  .\label{DeltaEp}%
\end{equation}
The localization of the impurities comes at a kinetic energy cost and the
impurity-impurity overlap comes at an interaction energy cost. We describe the
direct impurity-impurity interactions by an effective potential $v_{II}\left(
\mathbf{r}_{i}-\mathbf{r}_{j}\right)  =\lambda_{II}\delta\left(
\mathbf{r}_{i}-\mathbf{r}_{j}\right)  $. Adding $N$ strongly coupled impurities 
to a homogeneous dilute BEC changes the ground state energy by an amount $E$ where
\begin{eqnarray}
E  & =N\lambda_{IB}\rho_{0}\nonumber\\
& +\int d^{3}\mathbf{r}_{1}\int d^{3}\mathbf{r}_{2}\ldots\int d^{3}%
\mathbf{r}_{N}\ \psi^{\ast}\left(  \mathbf{r}_{1},\mathbf{r}_{2}%
,\ldots\mathbf{r}_{N}\right)  \sum_{j}\left(  \frac{-\hbar^{2}\nabla_{j}^{2}%
}{2m_{I}}\right)  \psi\left(  \mathbf{r}_{1},\mathbf{r}_{2},\ldots
\mathbf{r}_{N}\right)  \nonumber\\
& +\Delta E_{p}\nonumber\\
& +\int d^{3}\mathbf{r}_{1}\int d^{3}\mathbf{r}_{2}\ldots\int d^{3}%
\mathbf{r}_{N}\;\int d^{3}\mathbf{r}_{1}^{\prime}\int d^{3}\mathbf{r}%
_{2}^{\prime}\ldots\int d^{3}\mathbf{r}_{N}^{\prime}\;\psi^{\ast}\left(
\mathbf{r}_{1},\mathbf{r}_{2},\ldots\mathbf{r}_{N}\right)  \nonumber\\
& \times\frac{1}{2}\sum_{i\neq j}v_{II}\left(  \mathbf{r}_{i}-\mathbf{r}%
_{j}^{\prime}\right)  \psi\left(  \mathbf{r}_{1}^{\prime},\mathbf{r}%
_{2}^{\prime},\ldots\mathbf{r}_{N}^{\prime}\right).
\end{eqnarray}
The mediated inter-particle interaction potential, $V_{s}$, is attractive ($V_{s}<0$) and favors
impurity overlap. Maximal overlap is achieved when the $N$ bosons occupy the
same single particle orbital, $\psi\left(  \mathbf{r}_{1},\mathbf{r}%
_{2},\ldots\mathbf{r}_{N}\right)  =\psi\left(  \mathbf{r}_{1}\right)
\psi\left(  \mathbf{r}_{2}\right)  \ldots\psi\left(  \mathbf{r}_{N}\right)  $.
To emphasize that the boson impurity particles not only self-localize, but
occupy the same orbital (localizing in Hilbert space), we will refer to the
co-self-localization of the bosonic impurity atoms. In that case, $\rho
_{I}\left(  \mathbf{r}\right)  =\left\vert \psi\left(  \mathbf{r}\right)
\right\vert ^{2}$ and using equation (\ref{DeltaEp}) in the $E$ expression we
find
\begin{eqnarray}
E  &  =N\lambda_{IB}\rho_{0}+N\int d^{3}\mathbf{r}\psi^{\ast}\left(
\mathbf{r}\right)  \left(  \frac{-\hbar^{2}\nabla^{2}}{2m_{I}}\right)
\psi\left(  \mathbf{r}\right) \nonumber\\
&  +\frac{N\left(  N-1\right)  }{2}\lambda_{II}\int d^{3}\mathbf{r}\left\vert
\psi\left(  \mathbf{r}\right)  \right\vert ^{4}\nonumber\\
&  +\frac{N^{2}}{2}\int d^{3}\mathbf{r}\int d^{3}\mathbf{r}^{\prime}\left\vert
\psi\left(  \mathbf{r}\right)  \right\vert ^{2}V_{s}\left(  \mathbf{r}%
-\mathbf{r}^{\prime}\right)  \left\vert \psi\left(  \mathbf{r}^{\prime
}\right)  \right\vert ^{2},\label{Emanybody}%
\end{eqnarray}
where $\psi\left(  \mathbf{r}\right)  $ is assumed to be normalized. In
determining the optimal $\psi$-orbital, we introduce a Lagrangian multiplier
$N\mu$ to ensure $\int d^{3}\mathbf{r}\left\vert \psi\left(  \mathbf{r}%
\right)  \right\vert ^{2}=1$. Minimizing $E-N\mu\left(  \int d^{3}%
\mathbf{r}\left\vert \psi\left(  \mathbf{r}\right)  \right\vert ^{2}-1\right)
$ with respect to $\psi^{\ast}$ yields
\begin{eqnarray}
\mu_{I}\psi\left(  \mathbf{r}\right)   &  =\frac{-\hbar^{2}\nabla^{2}}{2m_{I}%
}\psi\left(  \mathbf{r}\right)  +\left(  N-1\right)  \lambda_{II}\left\vert
\psi\left(  \mathbf{r}\right)  \right\vert ^{2}\psi\left(  \mathbf{r}\right)
\nonumber\\
&  -NQ^{2}\int d^{3}\mathbf{r}^{\prime}\frac{e^{-\left\vert \mathbf{x-r}%
\right\vert /\xi}}{\left\vert \mathbf{r-r}^{\prime}\right\vert }\left\vert
\psi\left(  \mathbf{r}^{\prime}\right)  \right\vert ^{2}\psi\left(
\mathbf{r}\right)  \;,\label{muIchif}%
\end{eqnarray}
where $\mu_{I}=\mu-\lambda_{IB}\rho_{0}$. The initial $N$-factor in the
Lagrange multiplier ensures that $\mu_{I}$ becomes the impurity chemical
potential in the limit of large impurity numbers $N$. Solving the above
equation for $\mu_{I}$ by multiplying both sides by $\psi^{\ast}\left(
\mathbf{r}\right)  $ and integrating over $\mathbf{r}$, we obtain
\begin{eqnarray}
\mu_{I}  &  =\int d^{3}\mathbf{r}\psi^{\ast}\left(  \mathbf{r}\right)  \left(
\frac{-\hbar^{2}\nabla^{2}}{2m_{I}}\right)  \psi\left(  \mathbf{r}\right)
\nonumber\\
&  +\left(  N-1\right)  \lambda_{II}\int d^{3}\mathbf{r}\left\vert \psi\left(
\mathbf{r}\right)  \right\vert ^{4}\nonumber\\
&  -N\int d^{3}\mathbf{r}\int d^{3}\mathbf{r}^{\prime}\left\vert \psi\left(
\mathbf{r}^{\prime}\right)  \right\vert ^{2}\frac{Q^{2}e^{-\left\vert
\mathbf{x-r}\right\vert /\xi}}{\left\vert \mathbf{r-r}^{\prime}\right\vert
}\left\vert \psi\left(  \mathbf{r}^{\prime}\right)  \right\vert ^{2}%
.\label{muI}%
\end{eqnarray}
The physical $\mu_{I}$ value minimizes $E$ in equation (\ref{Emanybody}). For
negative values of $\mu_{I}$, $\psi$ becomes a localized wavefunction. At a
distance exceeding $\xi$ from where the impurity density becomes small, $\psi$
takes on the Yukawa profile of a bound state.

\section{Variational description of BEC-impurity
co-self-localization\label{Sec_analysis}}

Alternative to solving the non-linear Schrodinger equation of
equation (\ref{muIchif}), we can approximate the impurity orbital $\psi
(\mathbf{r})$ variationally. As in section \ref{Sec_LandauPekar}, we insert a
normalized Gaussian wavefunction, $\psi(\mathbf{r})=e^{-\mathbf{r}^{2}%
/2\sigma^{2}}\left(  \pi\sigma^{2}\right)  ^{-3/4}$, into the energy $E$ of
equation (\ref{Emanybody}) and we minimize $E$ with respect to the Gaussian width
$\sigma$. The treatment predicts self-localization of the $N$ co-locating
impurities if $\mu_{I}$ of equation (\ref{muI}), evaluated at $\sigma$ that
minimizes $E$, gives a negative value.

\subsection{The variational functionals}

In the scaling units, $E_{0},R_{0}$, introduced above to treat the single
impurity polaron problem, the $E$ and $\mu_{I}$ expressions take the form
\begin{equation}
\frac{E\left(  N\right)  }{E_{0}}=-\frac{N^{2}}{x}\sqrt{\frac{2}{\pi}}f\left(
\frac{x}{\xi}\right)  +N\frac{3}{2x^{2}}+\frac{N\left(  N-1\right)  }{2}%
\frac{\eta}{x^{3}},\label{varE}%
\end{equation}
and
\begin{equation}
\frac{\mu_{I}\left(  N\right)  }{E_{0}}=-\frac{2N}{x}\sqrt{\frac{2}{\pi}%
}f\left(  \frac{x}{\xi}\right)  +\frac{3}{2x^{2}}+\left(  N-1\right)
\frac{\eta}{x^{3}},\label{varm}%
\end{equation}
where $x=\sigma/R_{0}$ and where the impurity-impurity coupling was defined
above (see equation (\ref{eta})), $\eta=\left(  \lambda_{II}R_{0}^{-3}\right)
/\left(  \left[  2\pi\right]  ^{3/2}E_{0}\right)  $.

Two interesting and relevant limits can be understood analytically: the
`pinprick' limit in which the size of the impurity wavefunction shrinks well
below the coherence length of the BEC and the large impurity number limit in
which the impurity wavefunction extent is determined by the competition of the
BEC-mediated interaction and the impurity-impurity repulsion.

\subsection{The pinprick polaron limit}

\begin{figure}[pth]
\begin{center}
\includegraphics[width=3in] {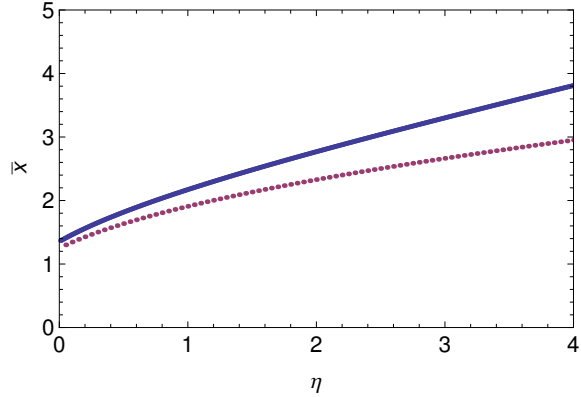}
\end{center}
\caption{The expected Gaussian width of the BEC impurity wavefunction that
describes the ground state of the BEC with three bosonic impurity atoms
occupying the same single particle (Gaussian) wavefunction, shown for
$\beta=4$ as a function of the impurity-impurity coupling constant $\eta$. The
upper curve, shown by the full line in blue, shows the variationally
determined width, whereas the lower graph (shown in red dots) pictures the
Gaussian width determined in the Taylor-expanded approximation. The Taylor
expansion treats the BEC mediated interaction as a Coulomb interaction and
underestimates the size of the impurity wavefunction, which grows as a
consequence of the short-range impurity-impurity repulsion.}%
\label{fig2}%
\end{figure}

The coherence length $\xi$ is the natural length scale of the BEC. If an
object has a subcoherence length size $L$, i.e., $L\ll\xi$, the BEC response
to its presence is indistinguishable from the BEC-response to a point object.
In the presence of $N$ co-self-localizing boson impurities, the BEC-mediated
impurity-impurity attractions that contribute a term $\sim N^{2}$ can further
decrease the impurity size to be sub-coherence length --- the impurity becomes
a `pinprick'. In this limit, the $f$ argument, $x/\beta=\sigma/\xi$, is much
less than $1$ and $f$ can be Taylor-expanded in the $E$ and $\mu_{I}$ expressions,
\begin{equation}
E\left(  x;N,\beta,\eta\right)  =E_{c}\left(  x;N,\eta\right)  +\frac{NE_{0}%
}{\beta},\label{vartE}%
\end{equation}
and
\begin{equation}
\mu_{I}\left(  x;N,\beta,\eta\right)  =\mu_{c}\left(  x;N,\eta\right)
+\frac{2NE_{0}}{\beta},\label{vartm}%
\end{equation}
where $E_{{c}}$ and $\mu_{c}$ denote the Gaussian expectation values for
Coulomb (instead of Yukawa) self-interactions,
\begin{eqnarray}
\frac{E\left(  N\right)  }{E_{0}}  & =-\frac{N^{2}}{x}\sqrt{\frac{2}{\pi}%
}+N\frac{3}{2x^{2}}+\frac{N\left(  N-1\right)  }{2}\frac{\eta}{x^{3}%
},\nonumber\\
\frac{\mu_{I}\left(  N\right)  }{E_{0}}  & =-\frac{2N}{x}\sqrt{\frac{2}{\pi}%
}+\frac{3}{2x^{2}}+\left(  N-1\right)  \frac{\eta}{x^{3}}.
\end{eqnarray}
In this limit, the equilibrium Gaussian width follows from the requirement
$\partial E_{c}/\partial x=0$ at fixed $N$ and $\eta$, yielding $\overline
{x}\left(  N,\eta\right)  $. The critical coupling $\beta_{c}\left(
N,\eta\right)  $ follows from $\mu_{I}=0$;
\begin{equation}
\beta_{c}\left(  N,\eta\right)  \simeq-\frac{2N}{\mu_{c}\left(  \bar{x}%
;N,\eta\right)  /E_{0}},\label{bcT}%
\end{equation}
although this approach becomes somewhat questionable for determining the
impurity properties near the critical coupling since the Taylor expansion is
not generally valid near the self-localization point. Furthermore, if the
short-range impurity-impurity interactions can be neglected ($\eta
\rightarrow0$) in the pinprick multi-impurity polaron, the condition,
$\partial E_{c}/\partial x=0$ leads to the equilibrium width $\overline
{x}=\left[  3\sqrt{\pi/2}\right]  /N\approx\overline{x}\left(  N=1,\eta
=0\right)  /N$. The expectation value of $E_{c}$ scales as $N^{3}$ and the
binding energy takes the form
\begin{equation}
E\left(  N;\beta,\eta=0\right)  =-\frac{N^{3}}{3\pi}E_{0}+\frac{N}{\beta}%
E_{0}.
\end{equation}
On the other hand, the expectation value of $\mu_{c}$ scales as $N^{2}$ ,
$\mu_{c}/E_{0}=-N^{2}/\pi$ and from equation (\ref{bcT}) we estimate the critical
coupling constant as
\begin{equation}
\beta_{c}\approx\frac{2\pi}{N}.
\end{equation}
Hence, $N$ bosonic impurities can co-self-localize at much smaller values of
the boson-impurity coupling strength than the critical coupling for
single-impurity self-localization. The self-localized multi-impurity polaron
can be significantly smaller than the BEC-coherence length. Specifically,
casting the impurity wavefunction extent, $\sqrt{\left\langle r^{2}%
\right\rangle }=\sqrt{3/2}\sigma=\sqrt{3/2}\overline{x}R_{0}$ and the binding
energy in terms of the BEC-length and energy scales, we find
\begin{eqnarray}
\sqrt{\left\langle r^{2}\right\rangle }  & =\frac{3\sqrt{3\pi}}{2\pi N}%
\frac{\xi}{\beta},\nonumber\\
E  & =2\left(  \frac{m_{B}}{m_{I}}\right)  \mu_{B}\left(  -\frac{N^{3}}{3\pi
}\beta^{2}+N\beta\right)  .
\end{eqnarray}
For small, but finite impurity-impurity coupling $\eta$, the impurity-impurity
interaction becomes significant as $N$ increases. Solving for the equilibrium
width $\overline{x}$ from $\partial E_{c}/\partial x=0$, we obtain
$\overline{x}=\left(  3\sqrt{\pi/2}\right)  \left(  1/2\right)  \left(
1+\sqrt{1+\left(  2/3\right)  \eta N(N-1)\sqrt{2/\pi}}\right)  $, so that the
impurity-impurity repulsion significantly increases the extent of the impurity
wavefunction when $N(N-1)\geq\left(  3\sqrt{\pi/2}/2\eta\right)  $. In
figure \ref{fig2} we show the Gaussian width of the three-impurity polaron
wavefunction at coupling $\beta=4$ as a function of $\eta$, as determined
numerically from the full functionals, equations (\ref{varE}) and (\ref{varm}), and
as determined from the Taylor expanded approximation equations (\ref{vartE}) and
(\ref{vartm}). 
Notice that the Taylor expansion only works well for small values
of $\eta$ as the impurity-impurity repulsion sufficiently increases the
impurity radius to invalidate the Taylor expansion assumption $\overline
{x}/\beta\ll1$. As the expansion replaces the Yukawa self-interaction by a
Coulomb attraction (more efficient at binding the polaron), the $E_{c}$
approximation underestimates the size of the impurity wavefunction.

\begin{figure}[pth]
\begin{center}
\includegraphics[width=3.3in] {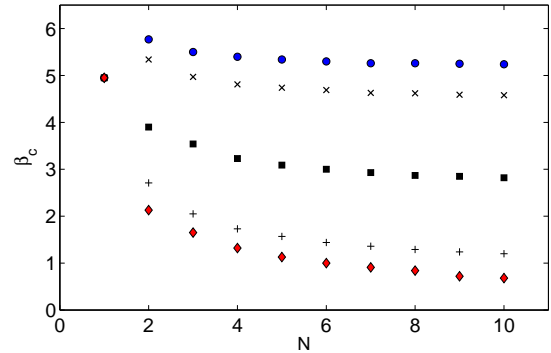}
\end{center}
\caption{Critical values of the impurity-boson coupling constant $\beta$,
$\beta_{c}$, above which N bosonic impurity atoms self-localize in the
BEC-impurity ground state described in the strong coupling approximation. The
graph shows $\beta_{c}$ for different impurity atom numbers $N$, $N=1,2,\ldots,10$
, set out on the horizontal axis, and gives the variationally
determined $\beta_{c}$-values for different impurity-impurity coupling values, $\eta$:
red diamonds-- $\eta=0$,
black plus-- $\eta=1$,
black square-- $\eta=10$,
black x-- $\eta=30$ and
blue circle-- $\eta=40$.}%
\label{fig3}%
\end{figure}

The effect of the impurity-impurity repulsion is particularly striking in the
dependence of the critical $\beta_{c}$ on the impurity numbers $N$, as shown
in figure \ref{fig3}. 
This figure shows $\beta_{c}(N,\eta)$ for $N=1$ (independent of
$\eta$), $2,3,...,10$, at $\eta=1,10,30$ and $40$. At $\eta=0$, $\beta
_{c}(N,\eta=0)$ decreases with increasing impurity numbers and $\beta
_{c}(N,\eta=0)$ vanishes at large $N$. At finite, fixed $\eta$ with $\eta<24$
the critical coupling $\beta_{c}(N,\eta)$ (shown by dots in the same color)
still decrease with increasing $N$, but $\beta_{c}(N,\eta)$ tends to a finite
value $\beta_{C}(N\rightarrow\infty,\eta)$ that depends on $\eta$. At
$\eta>24$, $\beta_{c}(2,\eta)>\beta_{c}(1)$, but then $\beta_{c}(N,\eta)$,
$N\geq2$ still falls off monotonically with increasing impurity numbers $N$.

This dependence has interesting implications. If the BEC-impurity system has
an impurity-boson coupling constant $\beta$ in the range $\left[  \beta
_{c}(N_{\beta}-1),\beta_{c}(N_{\beta})\right]  $, the above description
suggests that $N_{\beta}$ or more bosonic impurities would co-self-localize in
the ground state, whereas $N_{\beta}-1$ or less impurities would not. The
minimal number of particles required to make a stable, self-localized
structure is reminiscent of the familiar nucleation processes that take place
in first-order transitions. In that case, the formation of a stable droplet of
the new phase also requires a minimal amount of matter in the new phase to
overcome the relative cost of the surface energy (here, kinetic energy plays
the role of surface energy). We tentatively identify the co-self-localization
of impurities as the onset of the nucleation process of B-I phase separation,
a connection we will further explore below.

\subsection{BEC-phase separation}

In a phase separation transition, initially miscible fluids become immiscible
and separate from each other. At zero temperature, the ground state of a gas
of I and B weakly interacting boson particles changes from a mixed to a
separated configuration as the impurity-boson scattering length is increased.
The natural coupling constant to characterize this zero temperature phase
transition is given by
\begin{equation}
g=\frac{\lambda_{IB}^{2}}{\lambda_{II}\lambda_{BB}}=\frac{a_{IB}^{2}}%
{a_{II}a_{BB}}\frac{1}{4}\left(  1+\frac{m_{B}}{m_{I}}\right)  \left(
1+\frac{m_{I}}{m_{B}}\right)  .\label{g}%
\end{equation}
As $g$ varies from $g<1$ (miscible mixture of I- and B- BECs) to $g>1$ (ground
state corresponding to separated BECs), the system undergoes a phase
transition that appears to be second order - the compressibility and a
coherence length diverge. If the I BEC is the smaller condensate, the ground
state gathers the I-bosons into a spherical region as this geometry minimizes
the surface energy. The surface tension itself vanishes as $g\rightarrow1^{+}%
$, but is finite when $g>1$. Hence, we expect separation dynamics at $g>1$ to
be similar to that of a first-order transition.

For large I- and B- BECs, the onset of the separation dynamics is described by
the growing instability of the collective excitations of the homogeneous
BEC-mixture. But how does the separation dynamics set in if only a few I-atoms
are present and the I-bosons have not acquired macroscopic phase coherence,
whereas the temperature is sufficiently low to treat the B-BEC as a zero
temperature system? We suggest that when fluctuations gather a minimal number
$N_{\beta}$ of I-bosons in a B-coherence length size region, the impurities
can co-self-localize into a self-localized multi-impurity polaron (at least in
the parameter region in which the BEC density response to the impurities
remains small). Further collisions with more impurity atoms grow the
self-localized polaron impurity numbers. Finally, the relative B BEC density
response becomes large, the B-BEC density is expelled from the impurity region
and the impurity extent grows to exceed the B-coherence length. Below, we back
up this picture by showing that the condition for first order impurity
nucleation coincides (or nearly coincides) with the condition for zero
temperature BEC phase separation.

\subsection{The limit of large co-self-localizing impurity numbers}

As the impurity number $N$ increases at fixed $\beta$ coupling, the BEC
density response to the co-self-localizing impurity polaron grows. For an
impurity number $N_{D}$, the BEC density becomes comparable to the average
BEC-density and the self-localized impurity structure crosses over to a
`bubble' instead of a sub-coherence length polaron. In that regime, the
Frohlich-coupling of equation (\ref{Froh}), which was derived in the Bogoliubov
approximation and assumes low depletion, does not accurately describe the
impurity-boson coupling and the Landau-Pekar treatment breaks down.

Hence, the large impurity limit, $N\rightarrow\infty$, of $\beta_{c}$ further
strengthens the connection between multi-impurity polaron formation and BEC
phase separation. Keeping only the $N^{2}$-terms of the $E$ of
equation (\ref{varE}), the kinetic energy term drops out and the polaron size is
determined from the competition of the BEC B mediated interactions and the
short-range impurity-impurity interactions,
\begin{equation}
\frac{E_{N\rightarrow\infty}(N)}{E_{0}}=N^{2}\left[  \frac{\eta}{2x^{3}}%
-\sqrt{\frac{2}{\pi}}\frac{1}{x}f(\frac{x}{\beta})\right]  .
\end{equation}
where $f(x/\beta)=\int_{0}^{\infty}u\exp\left(  -\frac{u^{2}}{2}\right)
\exp\left(  -u\frac{x}{\beta}\right)  du$. Introducing $y=x/\beta$ as
variable, $y=\sigma/\xi$, the energy functional takes the form
\begin{equation}
\frac{E_{N\rightarrow\infty}(N)}{E_{0}}=A\left[  \frac{1}{y^{3}}-\kappa
\frac{f(y)}{y}\right]  ,
\end{equation}
where $A=N^{2}/\left[  2\beta^{2}\eta\right]  >0$ and $\kappa=\sqrt{8/\pi
}\beta^{2}\eta^{-1}$. This function has a minimum in $y$ and the local minimum
has a negative value provided $\kappa$ exceeds a minimal value $\kappa_{c}$.
Numerically, we found $\kappa_{c}=1.0004$. When $\kappa>\kappa_{c}$, the value
of $\mu_{I}$ at the minimum $E$\ is negative as $\mu_{I}=2E/N$ in the absence
of the kinetic energy term. Therefore, a large number $N$ of bosonic
impurities co-self-localize if $\kappa>\kappa_{c} $ or $\beta>\left(
\pi/8\right)  ^{1/4}\sqrt{\eta}\sqrt{\kappa_{c}}=\beta_{c}(N\rightarrow
\infty)$. We assume that the nucleation occurs when a number, any number, of
impurities can co-self-localize. As $\beta_{c}(N)$ decreases with $N$,
$\beta_{c}\left(  N\rightarrow\infty\right)  $ represents the minimal coupling
value at which the transition takes place. Therefore, a transition sets in if
\begin{equation}
\beta>\left(  \pi/8\right)  ^{1/4}\sqrt{\kappa_{c}}\sqrt{\eta}=\left(
16\pi\right)  ^{1/4}\sqrt{\kappa_{c}}\sqrt{\gamma\beta}\sqrt{\frac{a_{II}%
}{a_{BB}}},
\end{equation}
where we used $\eta=8\sqrt{2}\beta\gamma(a_{II}/a_{BB})$. As a consequence,
the nucleation condition takes the form
\begin{equation}
\beta>4\sqrt{\pi}\kappa_{c}\gamma\left(  \frac{a_{II}}{a_{BB}}\right)
\end{equation}
replacing $\beta=\sqrt{\pi}\sqrt{\rho_{0}a_{IB}^{4}/a_{BB}}\left(
1+m_{I}/m_{B}\right)  \left(  1+m_{B}/m_{I}\right)  $ and $\gamma=\sqrt
{\rho_{0}a_{BB}^{3}}$, we obtain
\begin{equation}
\frac{a_{IB}^{2}}{a_{II}a_{BB}}\frac{1}{4}\left(  1+\frac{m_{I}}{m_{B}%
}\right)  \left(  1+\frac{m_{B}}{m_{I}}\right)  >\kappa_{c},
\end{equation}
which is very close to the phase separation condition of equation (\ref{g}) as
$\kappa_{c}=1.0004$.

What is $N_{D}$ at which the self-localized polaron structure crosses over to
a phase separated bubble? The peak BEC density variation for a Gaussian
impurity wavefunction takes the form
\begin{equation}
\frac{\delta\rho_{B}(\mathbf{x}=0)}{\rho_{0}}=\frac{16}{\pi^{1/4}}\beta
^{3/2}\gamma^{1/2}\sqrt{\frac{m_{B}}{m_{I}}}\frac{N}{\overline{x}}f\left(
\frac{\overline{x}}{\sqrt{2}\beta}\right)  \;.
\end{equation}
In the pinprick regime, $\overline{x}/\beta\ll1$ with weak impurity-impurity
interactions, $N(N-1)\ll\left(  3\sqrt{\pi/2}/\left[  2\eta\right]  \right)  $
so that $\overline{x}=3\sqrt{\pi/2}/N$, we obtain
\begin{equation}
\frac{\delta\rho_{B}\left(  \mathbf{x}=0\right)  }{\rho_{0}}=-\frac{16}%
{3}\left(  \frac{4}{\pi^{3}}\right)  ^{1/4}\sqrt{\frac{m_{B}}{m_{I}}}%
\beta^{3/2}\gamma^{1/2}N^{2}\;.
\end{equation}
Estimating $N_{D}$ from $\left\vert \delta\rho_{B}\left(  \mathbf{x}=0\right)
/\rho_{0}\right\vert =1$, we find
\begin{equation}
N_{D}\approx\sqrt{\frac{3\pi^{3/4}}{16\sqrt{2}}}\left(  \frac{m_{I}}{m_{B}%
}\right)  ^{1/4}\frac{1}{\left(  \gamma\beta^{3}\right)  ^{1/4}}.
\end{equation}
The $N^{2}$--scaling of $\delta\rho_{B}/\rho_{0}$ may convey the message that
multi-impurity polarons with a finite number of impurities is unlikely.
However, at larger impurity number, the self-localization also sets in at a
smaller value of $N$. The relative BEC density variation near
self-localization with $\beta\approx\beta_{c}\approx\left(  2\pi\right)  /N$
is
\begin{equation}
\left\vert \frac{\delta\rho_{B}\left(  \mathbf{x}=0\right)  }{\rho_{0}%
}\right\vert _{N,\beta=\beta_{c}(N)}=\frac{64}{3}\pi^{3/4}\sqrt{m_{I}/m_{B}%
}\gamma^{1/2}N^{1/2}\;
\end{equation}
which scales as $\sqrt{N}$ and can be satisfied experimentally for $N\neq1$.

\subsection{ Self-localized impurity polaron fluid}

It is interesting to note that if $a_{II}$ is replaced by $\xi$ in the phase
separation coupling constant $g$, the expression goes over into the expression
for the impurity-boson coupling constant $\beta$,
\begin{equation}
g\frac{a_{II}}{\xi}=\beta\;.
\end{equation}
At phase separation, $g=1$ and $\beta=\beta_{c}(N\rightarrow\infty)$ so that,
up to a factor $\sqrt{\kappa_{c}/1}$
\begin{equation}
\frac{a_{II}}{\xi}=\beta_{c}(N\rightarrow\infty)\;.
\end{equation}
From figure \ref{fig3}, it is clear that when $\eta$ is sufficiently large to
make $\beta_{c}(N\rightarrow\infty)$ comparable to or greater than $5$,
(implying $a_{II}\gg\xi$), then $\beta_{c}(1)<\beta_{c}(N>1)$. What is the
ground state of an impurity-BEC system with multiple impurities and an
impurity-boson coupling constant $\beta$ with $\beta_{c}(1)<\beta<\beta
_{c}(N\rightarrow\infty)$? The conditions suggest that one impurity atom
self-localizes but that the co-self-localization of two or more impurities is
prevented by the short-range impurity-impurity repulsion. Each impurity could
self-localize, but how do the self-localized single impurity polarons arrange
themselves? As $\eta$ is large, two self-localized polarons repel each other
at small distances but attract each other at larger distances ranging up to
$\xi$. In classical particle systems, these are the conditions at which a
crystal can form or a liquid of self-determined density. Would the
self-localized polaron liquid form a self-localized polaron crystal, liquid or BEC?

This regime is, however, not easily realized experimentally. In the absence of
any resonances, the cold atom BEC coherence length generally exceeds the
scattering length by one or two orders of magnitude. The realization of a
self-localized single impurity polaron liquid that does not nucleate into a
phase separated BEC, would require the enhancement of both the $a_{IB}$ and
the $a_{II}$. If one Feshbach resonance enhances $a_{IB}$ to increase $\beta$,
the magnetic field is `used up' and $a_{II}$ cannot be varied magnetically. We
suggest that in two-dimensional systems confinement-induced resonances can be
used to enhance the other scattering length.

The confinement of atoms to one or two dimensions by squeezing the trap
potential in the other dimensions induces resonances in the atom-atom
interactions. These occur when the extent of the wavefunction in the
transverse (squeezed) direction becomes comparable to the scattering length in
three dimensions \cite{conf1}, \cite{conf2}. As we expect one and
two-dimensional BEC or quasi-BEC systems to undergo phase separation as well
as exhibit self-localization of impurity atoms, we suggest that a combination
of a magnetically controlled Feshbach resonance and a confinement-induced
resonance may access self-localized single impurity polaron fluids.

\section{Conclusions\label{Sec_conclusion}}

The experimental observation of multi-impurity self-localized polarons would
provide an interesting class of structures that have a subcoherence length
size. BEC experiments can address fundamental, strong coupling questions: Does
polaron self-localization give transition like behavior? What is the effective
mass of self-localized BEC impurity polarons? Can the impurity extent, as well
as the polaron size be measured directly by species specific dipole
potentials? In addition to fundamental many-body research the realization of
these structures may bring new methods to bear for manipulating and probing
BEC systems. Self-localized multi-impurity polarons can be transported by a
laser beam of moving but broad focus. Can the movable self-localized polaron
carry qubit information? The creation of shock wave analogues in BEC's
required density perturbations that were smaller than the BEC-coherence length
\cite{Lene}, can other phenomena be triggered such as superradiance in phonon
radiation? The realization of a Josephson junction in BEC's described by the
usual linear Josephson coupling Hamiltonian requires a constriction that is
smaller than the BEC-coherence length. Can a trapped self-localized impurity
provide such constriction?

If many impurity atoms are present in the trap, the self-localized impurity
polarons may represent transient structures. The self-localization dynamics,
we suggest, describes the onset of the phase separation nucleation process. On
the other hand, existing cold atom techniques can isolate the self-localized
impurity droplet with a controlled number of impurities. For instance, a
species specific optical lattice can contain a well determined number $N$ of
impurity atoms per site. An overlapping BEC does not feel the optical lattice
but can self-localize the $N$-impurities. As self-localization depends on the
number of co-located impurity atoms, this technique could provide another
sensitive method for counting atoms in each well.

In summary, we have shown that N bosonic neutral impurity atoms embedded in a
dilute gas BEC can self-localize into a collective strongly coupled polaron
state. The critical impurity-boson interaction strength for self-localization
can be significantly lower than the corresponding value for a single impurity
polaron. We identify the self-localized impurity polarons as the initial
droplets of the BEC phase separation nucleation process. In the limit of large
impurity-impurity interactions, there may be a strongly interacting regime in
which self-localized single impurity polarons form a stable self-localized
polaron liquid. We have discussed the experimental challenge of realizing
these systems in the polaron regime and not as phase separated bubbles.
Finally, we have speculated on possible applications of this new class of
sub-coherence length BEC structure.

\section{Acknowledgments}

ET's work was funded by the LDRD-program of Los Alamos National Laboratory.
Some of this work was performed at the Aspen Center of Physics.


\begin{thebibliography}{99}
\bibitem {Vbook}In condensed matter, self-localized polarons are typically
small - the polaron size is comparable to the lattice constant. Also, as the
electron mass is much smaller than that of the lattice ions, the time scales
for electron and phonon modes naturally separate. For an overview of
traditional polaron physics, see, for instance \cite{M72book}.
\bibitem {TCOKTD09N}In reference [16], the dependence of the polaron size on
on coupling strength of a $^{6}$Li impurity atom imbedded in a $^{23}$Na BEC
reverses itself and increases in a small coupling interval of the $\alpha$
coupling strength, $(2.7,3.2)$, near the value $\alpha=3.1$ for which the
strong coupling description predicts self-localization.
\end{thebibliography}
\end{document}